\documentclass[10pt,conference]{IEEEtran}
\IEEEoverridecommandlockouts
\usepackage{amsmath,amssymb,amsfonts}
\usepackage{textcomp}
\usepackage{xcolor}    
\usepackage{wasysym} 
\usepackage{amsthm}  
\usepackage{graphicx,times,cite, epsfig, array, mathrsfs}  
\usepackage{array}
\usepackage{multirow}  
\usepackage[caption=false,font=normalsize]{subfig}
\usepackage{textcomp}
\usepackage{stfloats}
\usepackage{url}
\usepackage{verbatim}
\usepackage{algorithm}
\usepackage[noend]{algorithmic}
\usepackage{caption}
\usepackage{threeparttable}
\usepackage{cases}

\usepackage{cite}

\def\BibTeX{{\rm B\kern-.05em{\sc i\kern-.025em b}\kern-.08em
    T\kern-.1667em\lower.7ex\hbox{E}\kern-.125emX}}

\newtheorem{theory}{Theorem}

\newtheorem{remark}{Remark}

\begin{document}

\title{
Throughput Optimization in Cache-aided Networks: An Opportunistic Probing and Scheduling Approach
\vspace{-1.3cm}
}

\author{ 
\IEEEauthorblockN{
Zhou~Zhang\IEEEauthorrefmark{1},
Saman~Atapattu\IEEEauthorrefmark{2},
Yizhu~Wang\IEEEauthorrefmark{1}, and Marco~Di~Renzo\IEEEauthorrefmark{3}}
 \IEEEauthorblockA{
\IEEEauthorrefmark{1}Tianjin Artificial Intelligence Innovation Center (TAIIC), Tianjin, China.  \\
 \IEEEauthorrefmark{2}School of Engineering, RMIT University, Melbourne, Victoria, Australia. \\
 \IEEEauthorrefmark{3}Université Paris-Saclay, 3 Rue Joliot Curie, 91190 Gif-sur-Yvette, France. \\
\IEEEauthorblockA{Email:
\IEEEauthorrefmark{1}\{zt.sy1986, wangyizhuj\}@163.com;\,
\IEEEauthorrefmark{2}saman.atapattu@rmit.edu.au;\,
\IEEEauthorrefmark{3}marco.di-renzo@universite-paris-saclay.fr
}
}
}

\maketitle

\begin{abstract}
This paper addresses the challenges of throughput optimization in wireless cache-aided cooperative networks. We propose an opportunistic cooperative probing and scheduling strategy for efficient content delivery. The strategy involves the base station probing the relaying channels and cache states of multiple cooperative nodes, thereby enabling opportunistic user scheduling for content delivery. Leveraging the theory of Sequentially Planned Decision (SPD) optimization, we dynamically formulate decisions on cooperative probing and stopping time. Our proposed Reward Expected Thresholds (RET)-based strategy optimizes opportunistic probing and scheduling. This approach significantly enhances system throughput by exploiting gains from local caching, cooperative transmission and time diversity. Simulations confirm the effectiveness and practicality of the proposed Media Access Control (MAC) strategy.
\end{abstract}

\begin{IEEEkeywords}
Opportunistic cooperative probing, Sequentially planned decision (SPD), Wireless cache-aided networks. 
\end{IEEEkeywords}

\section{Introduction} \label{s:intro}
\vspace{-0.1cm}
The surge in mobile data traffic, driven by content-centric services such as video streaming, app downloads/updates, and mobile TV, calls for effective data management strategies. With advanced generation and asynchronous access to content, {\it caching} at network edges, like base stations (BSs) and user devices, during low-load periods can significantly reduce delivery latency~\cite{yang2022}. Caching involves intermediate nodes prefetching content using various algorithms, reducing traffic and delay from the server to the end user. {\it Opportunistic scheduling}~\cite{Le2023}, when combined with caching-aided delivery, enhances spectrum efficiency in cache-aided cooperative networks. This integration has driven research into efficient caching and bandwidth resource utilization, leading to the development of joint cache-aided scheduling strategies. This paper focuses on exploring these strategies.


\vspace{-1mm}
\subsection{Related Work} 
\vspace{-1mm}

Probabilistic caching, particularly at intermediate nodes with limited storage, has been extensively studied to exploit the dynamic characteristics of network status~\cite{Ma2017,GaoJ2021,Junchao2019}.
In cellular systems, caching at BSs reduces core network traffic by employing optimized policies for maximizing throughput~\cite{Ma2017}. 
Device-to-Device (D2D) networks focus on optimal channel assignment strategies to minimize average content delivery delays~\cite{Junchao2019}, and optimal caching strategy is proposed for extensions with time-varying popularity~\cite{GaoJ2021}.
Recent studies have focused on cache-aided relaying transmission strategies for wireless content delivery~\cite{Zhou2015,dong2015,xie2023,Saputra2021,Jinsen2023,tang2023,Zhou2024,Zhou2024_tvt,Minh2024}. 
Studies~\cite{Zhou2015,dong2015,xie2023} find that cooperative transmission outperforms direct transmission in improving delivery outage performance when the caching node transmission fails. 
Another study~\cite{Saputra2021} proposes a joint caching and delivery strategy for ultra-high-rate low-latency communications, solving content delivery delay minimization as a mixed-integer nonlinear programming problem, and also devises an optimal cooperative node probing and scheduling strategy. 
In \cite{Jinsen2023}, collaborative caching and transmission power allocation 
are investigated for UAVs communications, and reinforcement learning based optimization algorithm is proposed to minimize delivery delay.
In \cite{tang2023,Zhou2024,Zhou2024_tvt}, multiple relays collaborate caching and multiple modes delivery is developed
to aid one source-to-destination transmission. Asymptotic and analytical expression for outage probability is derived, and caching strategies are optimized to improve the outage probability.
In addition, \cite{Minh2024} extends cooperative cache-aided scheduling techniques to satellite-UAV-terrestrial networks where 
the network latency is minimized.

\vspace{-1mm}
\subsection{Problem Statement and Contributions}
\vspace{-1mm}

Existing research in cache-aided cooperative networks often focuses on efficient delivery of single static user, with a primary design objective on optimizing outage probability and caching placement. Time-varying dynamics of different users content demands and wireless links and CSI overhead in utilizing diverse wireless links are neglected.
Efficient user scheduling approach faces challenges due to the presence of multiple caching nodes, the need for adaptive delivery modes, and the influence of time-varying channel gains. These factors necessitate the development of efficient strategies that can effectively 
balance the tradeoff between cache-aided delivery gain, CSI acquisition overhead and delivery time.

Motivated by the need to optimize throughput for randomly requested user scheduling, we address the Joint Cooperative Probing and User Scheduling (JCPUS) problem with three key contributions: i) we propose a novel opportunistic scheme that accounts for the dynamic nature of wireless networks, enabling more efficient content delivery; ii) we develop a new analytical framework that utilizes sequential decision optimization to adapt to changing network conditions in real-time; and iii) we present an SPD-optimized solution designed to maximize system throughput, thereby significantly enhancing overall network performance.

\vspace{-1mm}
\section{System Model}\label{s:system_model}
\vspace{-1mm}
 \begin{figure}[t!]
\begin{center}
\includegraphics[scale=.4]{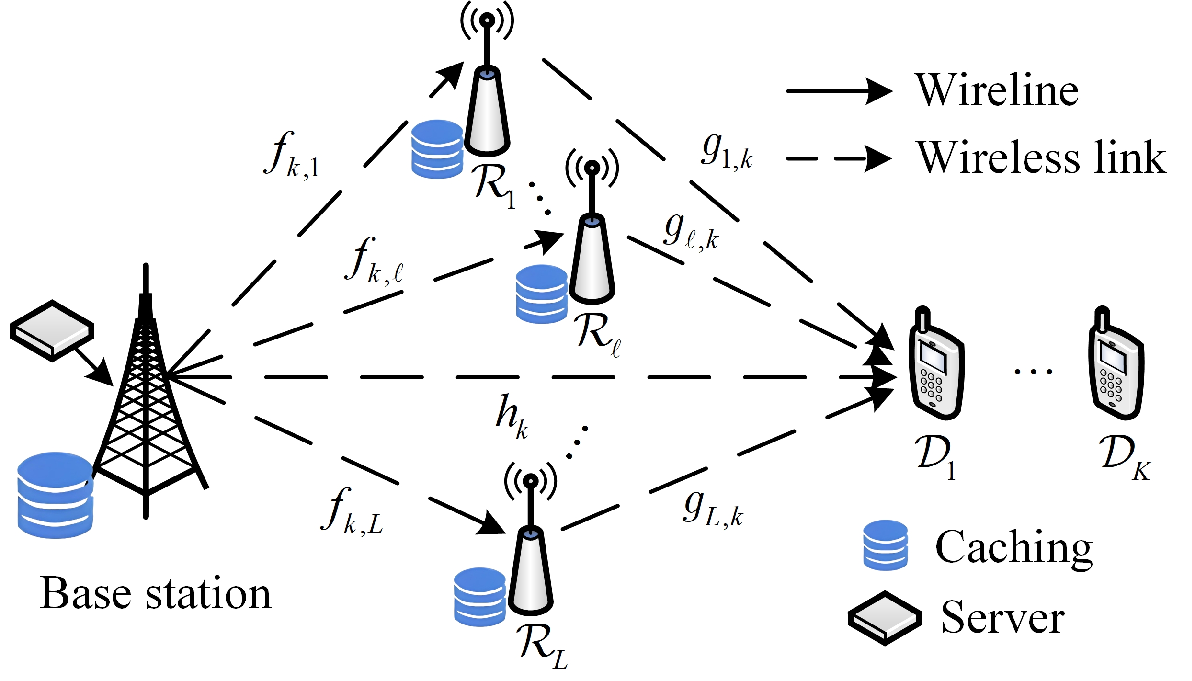}
\caption{{Illustration of a cache-aided cooperative network}}\label{f:system_mod2}
\end{center}
\vspace{-8mm}
\end{figure}
In Fig.~\ref{f:system_mod2}, we study a cache-aided cooperative network where users $\mathcal{D}_k, k\in\mathbb{N}$ request content from a server connected to the base station (BS) via wireline. The request time of users $\mathcal{D}_k$ is $t_{k}$, with $t_{0}=0$. The content request process follows a Poisson process with parameter $\lambda_d$, implying that the time interval $\tau_k=t_{k}-t_{k-1}$ follows independent exponentially distributed random variables with mean $\tau_s=1/\lambda_d$~\cite{hai2018}. 
There are $I$ contents requested by all users, with content size denoted as $q_i, i=1,\ldots,I$. The popularity of the requested contents follows the Zipf distribution, given by $p_i=\frac{i^{-\zeta}}{\sum_{u=1}^I u^{-\zeta}}$, where $\zeta$ is the skewness parameter.
To facilitate content delivery, multiple cooperative nodes $\mathcal{R}_{\ell}, \ell=1,...,L$ are deployed. Both the BS and cooperative nodes are equipped with caches of finite size. The caching probability of content $i$ at the BS and cooperative nodes is $p_{i}^b$ and $p_{i}^r$, respectively, satisfying $\sum_{i=1}^I p_{i}^b=C_b$ and $\sum_{i=1}^I p_{i}^r=C_r$, where $C_b$ and $C_r$ are the caching capacities of the BS and cooperative nodes.

\subsection{Wireless Channel Model}\label{sub:mac_channel}
For wireless channels, we consider a 2D geometric model with the BS fixed at the origin and cooperative nodes $\mathcal{R}_{\ell}, \ell=1,...,L$ at point $z_\ell$, respectively. Let $R$ be the BS coverage radius where users are uniformly distributed with probability density $1/\pi R^2$.
For user $\mathcal{D}_k$ at point $z=(z_X,z_Y)$, let $d_{k}$ denote the distance $||z||_2$ from the user to BS, $d_{1,\ell}$ denote the distance $||z_\ell||_2$ from the BS to cooperative node $\ell$ and $d_{2,\ell,k}$ denote the distance $||z-z_\ell||_2$ from cooperative node $\ell$ to the user. 
For wireless links, channel coefficients follow the Rayleigh fading. 
The received power for the link between BS and $\mathcal{D}_k$ is $P_t h_{k}$ with equivalent channel gain $h_{k}=d_{k}^{-2\alpha_1}\cdot\overline{h}_k$ and $\overline{h}_k\sim \exp(1)$; 
received power for the link from BS to $\mathcal{R}_{\ell}$ and for the link from $\mathcal{R}_{\ell}$ to $\mathcal{D}_k$ are 
${f}_{k,\ell}=d_{1,\ell}^{-2\alpha_2}\cdot \overline{f}_{k,\ell}$ and ${g}_{\ell,k}=d_{2,\ell,k}^{-2\alpha_2}\cdot \overline{g}_{\ell,k}$ respectively, 
where $\overline{f}_{k,\ell},\overline{g}_{\ell,k}\sim\exp(1)$ independently.

\vspace{-1mm}


\subsection{Opportunistic Scheduling Scheme}\label{sub:mac_scheme}
We propose the Joint Cooperative Probing and User Scheduling (JCPUS) scheme, an opportunistic approach that leverages the sequential plan decision concept for efficient user scheduling of random content requests~\cite{Xu2021,Zhou2023}.

At the start of a content delivery time frame ($t_0=0$), the BS awaits content requests from users. The process unfolds as follows:
i) After a random time interval $\tau_k$, User $\mathcal{D}_k$ transmits its content request $n_k\in\{1,..,I\}$ with pilot signals to the BS. The BS then acquires the file size $Q_k$, its caching state $\beta_{b,k}\in\{0,1\}$, and instantaneous channel gain $h_{k}$; 
ii) Cooperative nodes receive the request, obtaining their caching state $\beta_{r,\ell,k}\in\{0,1\}$ and instantaneous channel gain $g_{\ell,k}$;
iii) The BS calculates the instantaneous reward $\Lambda_1$ for direct delivery and the expected reward $\Lambda_2$ for cache-aided delivery based on random variables $(Q_k,\beta_{b,k},h_{k})$. It then compares the achievable rewards $\Lambda_1$ and $\Lambda_2$ with the threshold $\Lambda_3$ and chooses one of three options. The design parameters are discussed in Section~\ref{s:optimal-DOCA-stratg}.
\begin{enumerate}
    \item {\tt Direct delivery}: If $\Lambda_1 \ge \max\{\Lambda_2,\Lambda_3\}$, the BS schedules $\mathcal{D}_k$ for direct content delivery during $T_{k,1}$ (Section \ref{sub:relay_scheme}).
    \item {\tt Drop:} If $\max\{\Lambda_1,\Lambda_2\}<\Lambda_3$, the BS drops $\mathcal{D}_k$ and waits for the next user $\mathcal{D}_{k+1}$.
    \item {\tt Cooperative probe:} If $\Lambda_2>\max\{\Lambda_1,\Lambda_3\}$, the BS probes cache-aided nodes, determining the number of nodes ($J$). For $J\le L$, it randomly selects $J$ relays from set ${\mathcal L}_k$, which sequentially feedback channel gain $g_{\ell,k}$ and storage status $\beta_{r,\ell,k}\in\{0,1\}$ over duration $J\tau$. The BS then calculates the instantaneous reward $\Gamma_1$ based on $\beta_{r,\ell,k}$ and $g_{\ell,k}$, and decides on delivery or not by comparing $\Gamma_1$ with threshold $\Gamma_2$ (Section~\ref{s:optimal-DOCA-stratg}). 
    \begin{enumerate}
        \item {\tt Cache-aided delivery}: If $\Gamma_1\ge \Gamma_2$, the BS schedules $\mathcal{D}_k$ for cache-aided content delivery (Section \ref{sub:relay_scheme}).
        \item {\tt Drop:} If $\Gamma_1<\Gamma_2$, the BS drops $\mathcal{D}_k$ and waits for the next user.
    \end{enumerate}
\end{enumerate}
Upon scheduling a user, a new time frame commences.

\vspace{-1mm}
\subsection{Optimal Delivery Modes}\label{sub:relay_scheme}
This section outlines the optimal delivery modes and latency for both direct and cooperative delivery.

\subsubsection{Direct delivery}
For direct delivery from the BS to user $\mathcal{D}_k$, the delivery mode is optimized based on $\beta_{b,k}=0,1$. The delivery latency is given by $t_{k,1}=\frac{Q_k}{\log_2(1+P_t h_{k})}+\mathbb{I}[\beta_{b,k}=0]\cdot T_1$, where $T_1$ is the additional time spent by the BS to fetch the file from the server before delivery.

\subsubsection{Cache-aided delivery}
For cache-aided delivery from the BS to user $\mathcal{D}_k$, the delivery mode is optimized based on $\beta_{r,\ell,k}=0,1, \forall \ell\in {\mathcal L}_k$ and channel gains $f_{k,\ell},g_{\ell,k}$ for $\ell\in{\mathcal L}_k$. The delivery latency is expressed as follows
\begin{align*}
t_{k,2}=&\min\Big\{
\frac{2 Q_k}{\log_2\big(1+P_t h_{k} +
P_t\max\limits_{\ell\in {\mathcal L}_k} \big\{\min\{f_{k,\ell},g_{\ell,k}\}\big\}\big)},
 \\&
\frac{Q_k}{\log_2\big(1+P_t h_k+P_t\sum\limits_{\beta_{r,\ell,k}=1,\ell\in{\mathcal L}_k} g_{\ell,k}\big)}\Big\}.  
\end{align*}
The first term represents the latency by {\it mode I delivery}, where the BS delivers to $\mathcal{D}_k$ via the best decode-and-forward (DF) relay among probed nodes in ${\mathcal L}_k$. The second term represents the latency by {\it mode II delivery}, where the BS and nodes that have stored the file cooperatively deliver to $\mathcal{D}_k$.


\section{Problem Formulation}\label{s:homo}
We apply the SPD theory to formulate the optimization problem for the JCPUS strategy, aiming to maximize the average system throughput of content delivery.

\subsection{Preliminaries of SPD Theory}
The SPD theory strategically uses observed information to achieve predetermined goals, addressing the challenges in selecting the optimal observation path and time. The SPD problem is characterized by:
\begin{itemize}
    \item A sequence of random variable (RV) sets: $\{\mathbf{X}_1,...,\mathbf{X}_n,...\}$, where $\mathbf{X}_n=\{X_{n,1},...,X_{n,|A_n|}\}$ comprises a set $A_n$ of RVs with known distributions.
    \item A feasible observation path (OP): $\mathbb{A}=\big\{(a_1, \ldots, a_n): n\in{\mathbb{N}}\big\}$, where $a_n\subset A_n$ represents a subset of observations.
    \item A sequence of reward functions: $\{z_1(\mathbf{x}_{1,a_1}),\ldots,$ $z_{\infty}(\mathbf{x}_{1,a_1},\mathbf{x}_{2,a_2},...)\}$ denotes the sequence of rewards associated with the observed RVs in the feasible paths.
\end{itemize}
In the optimization problem, the decision-maker strategically selects sets of random variables to observe at each time step. Based on the observed information, the decision-maker decides whether to stop or continue, aiming to maximize the expected reward. This involves choosing the optimal OP and time to stop, denoted as $N=(a_1, a_2, \ldots)$, to maximize $\mathbb{E}[Z_N]$, where $Z_N=z_N(\mathbf{X}_{1,a_1},\ldots,\mathbf{X}_{|N|,a_{|N|}})$ represents the random reward upon stopping.


To solve the optimization problem, we define the maximum attainable reward function. For an OP $\mathbf{a}=(a_1,...,a_n)$, the historical information is $\mathscr{B}_{\mathbf{a}}$. A prolonged path is denoted as $(\mathbf{a},A)$ for a set $A$. We introduce the reward expectation functions $U_{\mathbf{a}} {=} \sup\limits_{\mathbf{b}\ge {\mathbf{a}}}\mathbb{E}[Z_{\mathbf{b}}|\mathscr{B}_{\mathbf{a}}]$ and $V_{\mathbf{a}}{=}\sup\limits_{A\subset A_{n+1}}\mathbb{E}[U_{(\mathbf{a},A)}|\mathscr{B}_{\mathbf{a}}]$. 
The relation between two paths $\mathbf{b}\ge\mathbf{a}$ implies $b_i=a_i$ for all $i\in \{1,2,\ldots, n\}$. For OP $\mathbf{a}$, $U_{\mathbf{a}}$ and $V_{\mathbf{a}}$ represent the expected maximum rewards with and without the {\it stop} decision, respectively. 
For a history with $|\mathbf{a}|=0$ (i.e. not making any observation yet), the expected reward $U_{\mathbf{a}}$ is denoted as $U_0$ such that $U_0=\sup\limits_{N}\mathbb{E}[Z_N]$. 
For OP ${\mathbf a}=(a_1,...,a_n)$, $V_{\mathbf a}$ and $U_{{\mathbf a}}$ are calculated from the Bellman Equation~\cite{Fenoy2017}. In particular, thresholds $V_{\mathbf a}$ and $U_{{\mathbf a}}$ satisfy that
\begin{equation}\label{e:bellman_equ}
U_{\mathbf a}\!=\!\max\{Z_{\mathbf a},V_{\mathbf a}\}\!=\!\max\big\{Z_{\mathbf a},\sup\limits_{A\subset A_{n+1}}\mathbb{E}[U_{(\mathbf{a},A)}|\mathscr{B}_{\mathbf{a}}]\big\}.
\end{equation} 
Theorem~\ref{th:optimal_rule} presents an optimal SPD rule based on these reward expectations.

\begin{theory}\label{th:optimal_rule} 
A Reward Expectation Threshold (RET) based rule, denoted as $N_1$, is defined as follows: Starting from 
$n=0$, at time step $n$, the rule decides to {\it stop} with $N_1=\mathbf{a}$ if $Z_{\mathbf{a}}\ge V_{\mathbf{a}}$, and {\it continue} otherwise. To {\it continue}, the OP is extended to $\mathbf{a}=(\mathbf{a},a_{n+1}^*)$, where $a_{n+1}^*=\min\{a_{n+1}\subset A_{n+1}:U_{(\mathbf{a},a_{n+1})}\!=\!V_{\mathbf{a}}\}$. 
If $\mathbb{E}\big[\sup\limits_{{\mathbf a}\in{\mathbb{A}}}Z_{{\mathbf a}}\big]<+\infty$, 
the RET-based rule $N_1$ is optimal, i.e. $N^*=N_1$, achieving $U_0=\sup\limits_{N} \mathbb{E}[Z_N]$. 
\end{theory}
\begin{IEEEproof}
It can be derived by Theorem 2.14 in \cite{Schmitz_N_book}.
\end{IEEEproof}



\subsection{SPD-Based Problem Formulation}\label{s:homo2}

We maximize system throughput by optimizing content observation, caching states, and channel conditions within an opportunistic delivery framework using SPD theory.

\subsubsection{Decisions Equivalence}
\label{sub:dec_transfer}
As outlined in Section~\ref{sub:mac_scheme}, following each content request, the choices are \texttt{direct delivery}, \texttt{drop}, and \texttt{cooperative probe}. Selecting \texttt{cooperative probe} specifies a node count $J \in \{1, \ldots, L\}$. We integrate the RET-based SPD rule from Theorem~\ref{th:optimal_rule} into our MAC framework by aligning both \texttt{drop} and \texttt{cooperative probe} under the SPD decision ``continue,'' albeit with varying subsequent observation sets. Specifically, if \texttt{drop} is selected, the next observation's reward is adjusted to be minimal to guarantee the BS omits the delivery.

\subsubsection{Sequential Observation Process}

In the JCPUS scheme, the BS acts as the decision-maker, initiating from time step $n = 1$ with content requests from users. An \textit{observation} refers to the acquisition of relevant random variables. For each user $\mathcal{D}_k$, the observation (Obs.) process includes:
\begin{itemize}
    \item First, the BS observes $(Q_k, \beta_{b,k}, h_{k})$ after a random interval $\tau_k$, capturing the information $\mathscr{F}_k = \{\tau_k, Q_k, \beta_{b,k}, h_k\}$ and makes a decision at step $(2k-1)$.
    \item If the decision is to \textit{continue}, the BS then observes RVs for $J$ selected nodes in $\mathcal{L}_k$ (for \texttt{drop}, $J=0$), where it captures $\mathscr{G}_k(J) = \{\beta_{r,\ell,k}, f_{k,\ell}, g_{\ell,k}, \ell \in \mathcal{L}_k\}$.
\end{itemize}
Each request logs two numbers for every step: $a_{2k-1} = 1$, ensuring $\mathscr{F}_k$ is observed, and $a_{2k}$, indicating the number of nodes observed in the second step, represented as $\mathscr{G}_k(a_{2k})$. To optimize rewards, decisions are based on prior observations up to a decision to \textit{stop}. The observation path up to time $n$ is denoted as $\mathbf{a} = (a_1, \ldots, a_n)$, and feasible observation paths are defined by the set $\mathbb{A} = \{(a_1, \ldots, a_n) : n \in \mathbb{N}, a_{2k-1} = 1, a_{2k} \in \{0, 1, \ldots, L\}, \forall k \leq \lceil n/2 \rceil\}$.

By time step $n$, the observed information under the observation path $\mathbf{a} = (a_1, \ldots, a_n)$ is $\mathscr{B}_{\mathbf{a}}$. For $n = 2k-1$, this information includes  $\mathscr{B}_{\mathbf{a}}=(\lor_{j=1}^k \mathscr{F}_j)\lor\big(\lor_{j=1}^{k\!-\!1} \mathscr{G}_k(a_{2k})\big)$; and for $n=2k$, the information is $\mathscr{B}_{\mathbf{a}}=(\lor_{j=1}^k\mathscr{F}_j)\lor\big(\lor_{j=1}^{k} \mathscr{G}_j(a_{2j})\big)$. The symbol $\lor_{j=1}^n$  denotes the union of information sets.

\subsubsection{Reward and Cost Functions}
The reward at any time step $n$, denoted $Y_{\mathbf{a}} = Q_k$, is defined as the content size delivered upon a \textit{stop} decision. The associated time cost, $T_{\mathbf{a}}$, accounts for the total duration from Obs. $1$ until Obs. $n$ plus the delivery latency. 
Specially,
at step $n=2k-1$, $T_{\mathbf a}=\sum_{j=1}^k {\tau}_j+\sum_{j=1}^{k\!-\!1}
\mathbb{I}_{[a_{2j}>0]} a_{2j}\tau
+t_{k,1}+\mathbb{I}[\beta_{b,k}=0]\cdot T_1$;
and at step $n=2k$, $T_{\mathbf a}=\sum_{j=1}^k {\tau}_j+\sum_{j=1}^{k}
\mathbb{I}_{[a_{2j}>0]} a_{2j}\tau+t_{k,2}$.

\subsubsection{Optimization Goal}
Let $N$ denote the sequential OP upon a {\it stop}. Here, $Y_N$ represents the reward function and $T_N$ the time cost at each stop by strategy $N$. In the JCPUS strategy for a single content delivery, the system throughput, defined as the ratio of the average reward to the average time cost, is $\mathbb{E}[Y_N]/\mathbb{E}[T_N]$ [bits/s]. The goal is to find an optimal strategy $N^*$ that maximizes the average system throughput $\eta^*$:
\begin{equation}\label{equ:stat_opt}
	N^*=\arg\sup\limits_{N}\frac{\mathbb{E}[Y_N]}{\mathbb{E}[T_N]}~~\text { and } ~~\eta^*=\frac{\mathbb{E}[Y_{N^*}]}{\mathbb{E}[T_{N^*}]}.
\end{equation}

\vspace{-2mm}


\section{Optimal JCPUS Strategy}\label{s:optimal-DOCA-stratg}
We now derive the optimal JCPUS strategy $N^*$ maximizing average system throughput based on the RET-based SPD rule.
\vspace{-2mm}
\subsection{Equivalent Ratio Optimization Problem}\label{sub:equal_transfer}
To optimize the throughput ratio, $\sup\limits_{N} \mathbb{E}[Y_N]/\mathbb{E}[T_N]$, we link it to a price-based objective. By introducing $\eta$ as the price on time cost, we define utility functions $Z_{\mathbf{a}}(\eta) = Y_{\mathbf{a}} - \eta T_{\mathbf{a}}$ and $Z_N(\eta) = Y_N - \eta T_N$. For a given $\eta > 0$, the rule achieving $\sup\limits_{N} \mathbb{E}[Z_N(\eta)]$ is denoted as $N(\eta)$, with an optimal rule represented by $N^*(\eta)$, which can be expressed as 
\begin{equation} 
	N^*(\eta)=\arg\sup\limits_{N}Z_N(\eta)=\arg\sup\limits_{N}\{Y_{N}-\eta T_{N}\}.
\end{equation}
The strategy $N^*(\eta^*)$ serves as the optimal $N^*$ for the ratio optimization problem, where $\eta^*$ is the unique value satisfying $\sup\limits_{N} \mathbb{E}[Z_N(\eta^*)] = 0$, and is given by $\eta^* = \sup\limits_{N} \mathbb{E}[Y_N] / \mathbb{E}[T_N]$.
\begin{figure*}[t!]
   \begin{align}
\Omega(\eta):=\sum\limits_{i=1}^I p_i \sum\limits_{\beta=0}^1 p_{i,\beta}^b \int\limits_{r=0}^R\int\limits_{\theta=0}^{2\pi}\int\limits_{x=0}^\infty\frac{r}{\pi R^2}
\max\!\big\{q_i-\eta (t_{k,1}+\beta T_1),0,
\max\limits_{\ell=1,\ldots,L}\!\!M_\ell(||z-z_\ell||_2^{-2\alpha_1}x,\beta,q_i,\eta)\big\} {\rm d}F_s(x) {\rm d}\theta {\rm d}r 
\label{e:optimality_equation2}
\end{align} \vspace{-2mm}\hrule\vspace{-7mm}
\end{figure*}
\subsection{Optimal Strategy}
We define reward function $M_\ell\big(h_s,\beta,q,\eta\big)$ ($\ell=1,\ldots,L$) as the expected reward if $\ell$ cooperative nodes are probed for 
user $\mathcal{D}_k$, expressed as $M_\ell\big(h_s,\beta,q,\eta\big):=
\mathbb{E}\big[\max\big\{q-\eta t_{k,2},0\big\}\big|h_k=h_s,\beta_{b,k}=\beta,Q_k=q\big]-\eta\ell\tau$.

\begin{theory}\label{th:optimalrule4}
The optimal JCPUS strategy $N^*$ achieving $\sup\limits_{N}\frac{\mathbb{E}[Y_N]}{\mathbb{E}[T_N]}$
is as follows: starting from $k=1$,
for user $\mathcal{D}_k$, 
the BS obtains $(Q_k,\beta_{b,k},h_k)$,
\begin{enumerate}
\item  if the immediate reward $ Q_k-\eta^* (t_{k,1}+\mathbb{I}[\beta_{b,k}=0]\cdot T_1) \ge \max\big\{\max\limits_{\ell=1,\ldots,L}M_\ell\big(h_k,\beta_{b,k},Q_k,\eta^*\big),0\big\}$, 
BS schedules $\mathcal{D}_k$ by {\tt direct delivery}; 
\item if the immediate reward 
$\max\big\{Q_k-\eta^* (t_{k,1}+\mathbb{I}[\beta_{b,k}=0]\cdot T_1),\max\limits_{\ell=1,...,L}M_\ell(h_k,\beta_{b,k},Q_k,\eta^*)\big\}<0$,
BS does not schedule by {\tt drop}.
\item otherwise, {\tt cooperative probe}.
Subsequently, 
the BS probes set ${\mathcal L}_k$ of $J^*$ cooperative nodes with $J^*=\min\big\{\ell\in\{1,\ldots,L\}:M_\ell(h_k,\beta_{b,k},Q_k,\eta^*)=
\max\limits_{u=1,\ldots,L}M_u(h_k,\beta_{b,k},Q_k,\eta^*)\big\}$. 
Then, after $J^*$ nodes are probed, BS obtains $(f_{k,\ell},g_{\ell,k},\beta_{r,\ell,k}),\ell\in{\mathcal L}_k$, calculates $r_1=\frac{1}{2}\log_2\big(1\!+\!P_t h_k\!+\!P_t\max\limits_{\ell\in {\mathcal L}_k} \big\{\min\{\!f_{k,\ell},g_{\ell,k}\!\}\big\}\big)$ and
$r_2=\log_2\big(1+P_t h_k +P_t\sum\limits_{\beta_{r,\ell,k}=1}g_{\ell,k}\big)\}$.
\begin{enumerate}
	\item if 
 $ \max\{r_1,r_2\}\ge \eta^*$, then {\tt cache-aided delivery};
 \begin{enumerate}
     \item 
     if 
     $r_1\ge r_2$, use {\it mode I delivery};
     \item otherwise, use {\it mode II delivery};
 \end{enumerate}
	\item otherwise, {\tt drop} by waiting until next user $\mathcal{D}_{k+1}$.
 \end{enumerate}
\end{enumerate}
The maximal throughput $\eta^*$ is uniquely determined by 
$\Omega(\eta^*)=\eta^*{\tau_s}$, where $\Omega(\eta)$ is presented in (\ref{e:optimality_equation2}) and the cumulative distribution function $F_s(x)$ for the channel gain $h_{k}$ with an exponential distribution. 
\end{theory}

\begin{IEEEproof}
 Utilizing the equivalence transfer method outlined in Section~IV-A, we address the maximization problem of system throughput through a three-step approach:

\paragraph{Refine the optimal rule $N^*(\eta)$ using the statistical property of the JCPUS problem} 

In accordance with the theoretical framework in Section~III-B, we derive the optimal SPD rule $N^*(\eta)$ as follows.
By definitions of $Y_{{\mathbf a}}$ and $T_{{\mathbf a}}$, we prove 
$  \mathbb{E}\big[\sup\limits_{{\mathbf a}\in{\mathbb{A}}}Z_{{\mathbf a}}(\eta)\big]
\!\le \mathbb{E}\big[Q_k\big]=\sum_{i=1}^I p_i q_i <+\infty$.

Then, 
based on Theorem~1, we transform the RET-based rule $N_1$ into SPD rule $N^*(\eta)$,
and its reward functions $U_{\mathbf a}(\eta)$ and $V_{\mathbf a}(\eta)$ are determined by solving Bellman equations:
\vspace{-1mm}
\begin{equation}\label{e:Aaexp}
\left\{
\begin{array}{ll}
U_{\mathbf a}(\eta)=\max\big\{Z_{\mathbf a}(\eta),V_{\mathbf{a}}(\eta)\big\}, \\
V_{\mathbf{a}}(\eta)=\max\limits_{\ell\in\{0,1,...,L\}}\mathbb{E}[U_{(\mathbf{a},\ell)}(\eta)|\mathscr{B}_{\mathbf{a}}],
~~~\text{Odd $n$};\\
U_{\mathbf a}(\eta)=\max\big\{Z_{\mathbf a}(\eta),V_{\mathbf{a}}(\eta)\big\}, \\
V_{\mathbf{a}}(\eta)=\mathbb{E}[U_{(\mathbf{a},1)}(\eta)|\mathscr{B}_{\mathbf{a}}],
~~~~~~~~~~~~~~~\text{Even $n$}.
\end{array} \right.
\end{equation}
\vspace{-1mm}


Given that each user request corresponds to two time steps, for an even time step $n=2k$, when $a_{2k}>0$, expression~(\ref{e:Aaexp}) transforms into
$U_{\mathbf a}(\eta) = \max\big\{Q_k-\eta t_{k,2},
U_0\big\}
 -\eta\big(T_c(k)+ a_{2k}\tau\big)$; 
when $a_{2k}=0$, the utility function $Z_{\mathbf a}$ is unbounded, i.e., $Z_{\mathbf a}=-\infty$. Consequently, expression (\ref{e:Aaexp}) takes on a specific form as:
\begin{equation}
\vspace{-1mm}
U_{\mathbf a}(\eta)=\mathbb{E}[U_{(\mathbf{a},1)}(\eta)|\mathscr{B}_{\mathbf{a}}]=U_0\!-\!\eta T_c(k).\label{e:iterate3}
\end{equation}
For odd step $n=2k-1$, by substituting (6) into (\ref{e:Aaexp}), we obtain
\vspace{-1mm}
\begin{align}\label{e:expres2}
U_{\mathbf a}(\eta)=& \max\big\{Q_k-\eta (t_{k,1}+\mathbb{I}[\beta_{b,k}=0] \cdot T_1),
U_0,\nonumber\\ & \max\limits_{\ell=1,\ldots,L}\overline{M}_\ell\big(h_k,\beta_{b,k},Q_k,\eta,U_0\big)\big\} 
\!-\!\eta T_c(k),
\end{align}
\vspace{-1mm}
where the reward function $\overline{M}_\ell\big(h_k,\beta_{b,k},Q_k,\eta,U_0\big)$ (for $\ell=1,\ldots,L$), representing the maximal average reward when $\ell$ cooperative nodes are probed for $\mathcal{D}_k$, is computed as:
\vspace{-1mm}
\begin{align}
    &\overline{M}_\ell\big(h_k,\beta_{b,k},Q_k,\eta,U_0\big) \nonumber\\ 
    &\qquad:=
\mathbb{E}\big[\max\big\{Q_k-\eta t_{k,2},U_0\big\}\big|h_k,\beta_{b,k},Q_k\big]-\eta\ell\tau.
\end{align}
Additionally, 
$V_{\mathbf a}$ can be determined using expression~(\ref{e:Aaexp}).



\paragraph{Transfer the SPD rule $N^*(\eta)$ to the JCPUS strategy} 


Utilizing the decision equivalence in Section~IV-A, we transition the SPD rule $N^*(\eta)$ to the JCPUS strategy and analyze the optimal decision conditions following each observation.

First, we examine the case when $n=2k\!-\!1$. The BS opts for {\tt direct delivery} when $Z_{\mathbf a}\ge V_{\mathbf a}$; otherwise, it prolongs the observation process by probing $J^*$ cooperative nodes, where $J^*=\min\big\{0\le \ell\le L:U_{({\mathbf a},\ell)}=V_{{\mathbf a}}\big\}$. The condition for {\tt direct delivery} can be expressed as:
\begin{align*}
&Q_k-\eta (t_{k,1}+\mathbb{I}[\beta_{b,k}=0]\cdot T_1)
\\&~~~~~~\ge \max\big\{\max\limits_{\ell=1,\ldots,L}\overline{M}_\ell\big(h_k,\beta_{b,k},Q_k,\eta,U_0\big),U_0\big\}.
\end{align*}


If the optimal decision is to \textit{continue} with $J^* > 0$, the optimal number of cooperative nodes to probe, denoted as $J^*$, is determined as the minimum value satisfying the condition:
\vspace{-1mm}
\begin{align*}
\overline{M}_{J^*}\!(h_k,\beta_{b,k},Q_k,\eta,U_0\!) \!=\! \!
\max\limits_{\ell\!=\!1,\ldots,L}\!\overline{M}_\ell(\!h_k,\beta_{b,k},Q_k,\eta,U_0\!)\! \ge\! U_0 
\end{align*}


If the optimal decision is to \textit{continue} with $J^* = 0$ (i.e., $V_{\mathbf{a}}=U_0-\eta T_c(k)$), the condition for {\tt drop} is determined as
\vspace{-1mm}
\begin{align*}
&\max\big\{Q_k-\eta (t_{k,1}+\mathbb{I}[\beta_{b,k}=0]\cdot T_1),
\\&~~~~~~~~~~~\max\limits_{\ell=1,\ldots,L}\overline{M}_\ell\big(h_k,\beta_{b,k},Q_k,\eta,U_0\big)\big\}
< U_0.
\end{align*}


Then, we examine the case when $n\!=\!2k$. If $a_n=J$, indicating that $J$ nodes are probed, the BS selects \texttt{cache-aided delivery} if $Z_{\mathbf a}\ge V_{\mathbf a}$. Otherwise, the BS opts for \texttt{drop}. Specifically, the condition for \texttt{cache-aided delivery} is:
\vspace{-2mm}
\begin{align*}
Z_{\mathbf a}=&
Q_k\!-\!\eta t_{k,2}
\!-\!\eta J \tau-\eta T_c(k)
\ge V_{\mathbf a}=
U_0\!-\!\eta J \tau-\eta T_c(k).
\end{align*}


Combining the aforementioned results, the optimal JCPUS strategy $N^*(\eta)$ for $\sup\limits_{N}\mathbb{E}[Z_N(\eta)]$, is as follows:
Starting from $k=1$, for user $\mathcal{D}_k$, the BS obtains 
$(Q_k,\beta_{b,k},h_k)$,
\begin{itemize}
	\item if the immediate reward $Q_k-\eta (t_{k,1}+\mathbb{I}[\beta_{b,k}=0]\cdot T_1)\ge \max\Big\{\max\limits_{\ell=1,...,L}\overline{M}_\ell\big(h_k,\beta_{b,k},Q_k,\eta,U_0\big),U_0\Big\}$, the BS schedules $\mathcal{D}_k$ through \texttt{direct delivery}.
	
	\item if $U_0> \max\big\{Q_k-\eta (t_{k,1}+\mathbb{I}[\beta_{b,k}=0]\cdot T_1), \max\limits_{\ell=1,...,L}\overline{M}_\ell\big(h_k,\beta_{b,k},Q_k,\eta,U_0\big)\big\}$, the BS does not schedule and opts for \texttt{drop}.
	
	\item otherwise, the BS engages in \texttt{cooperative probe} by probing $J^*$ cooperative nodes, where $J^*=\min\big\{\ell\in\{1,...,L\}:\overline{M}_\ell(h_k,\beta_{b,k},Q_k,\eta,U_0)=\max\limits_{u=1,\ldots,L}\overline{M}_u(h_k,\beta_{b,k},Q_k,\eta,U_0)\big\}$. Subsequently:
	\begin{itemize}
		\item if the immediate reward $Q_k-\eta t_{k,2}\ge U_0$, then \texttt{cache-aided delivery} is executed using the optimal delivery mode in Section II-B.
		\item otherwise, the BS chooses \texttt{drop}, and waits until the next user $\mathcal{D}_{k+1}$.
	\end{itemize}
\end{itemize}


The maximal expected reward $U_0$ is determined by 
\begin{align}\label{e:optimality_equation}
   U_0=&\mathbb{E}\big[\max\big\{Q_k-\eta (t_{k,1}+\mathbb{I}[\beta_{b,k}=0]\cdot T_1),U_0,
   \nonumber\\& ~~~~~~
\max\limits_{\ell=1,\ldots,L}\overline{M}_\ell(h_k,\beta_{b,k},Q_k,\eta,U_0)\big\}\big]-\eta\tau_s.
\end{align}
Leveraging the i.i.d. statistical property of RVs $(h_k,\beta_{b,k},Q_k)$ and interval time $\tau_k$ for $k\in\mathbb{N}$, it's noteworthy that for all $k\ge 1$, the right-hand side of (\ref{e:optimality_equation}) remains constant. 

\paragraph{Replace $\eta$ with $\eta^*$ and obtain optimal strategy $N^*$} 

By utilizing the equivalence transferring method and substituting $U_0$ and $\eta$ with $0$ and $\eta^*$, the optimal JCPUS strategy $N^*=N^*(\eta^*)$
is explicitly described in the theorem.
By calculating the expectation as in~\eqref{e:optimality_equation2} where $p_{i,1}^b=p_{i}^b, p_{i,0}^b=1-p_{i}^b, ||z-z_\ell||_2=||(r\cos\theta,r\sin\theta)- z_\ell||_2$,  $\eta^*$ can be obtained
as solution of $\Omega(\eta)=\eta\tau_s$.
In this context, $M_\ell(h_s,\beta_{b},q,\eta)$ is defined by evaluating $\overline{M}_\ell(h_s,\beta_{b},q,\eta,U_0)$ at $U_0=0$. 


\end{IEEEproof}
Based on Theorem~\ref{th:optimalrule4}, we have following two remarks. 
\vspace{-1mm}
\begin{remark}
The maximal throughput $\eta^*$ is determined solely by network statistics and is computed offline. The reward functions $M_\ell(h_k,\beta_{b,k},Q_k,\eta^*)$ for $\ell=1,\ldots,L$, which depend only on content size $Q_k$, channel gain $h_k$, and caching state $\beta_{b,k}$, can be efficiently calculated either online or pre-computed and stored in a lookup table. With the observed variables $(Q_k,\beta_{b,k},h_k)$, the optimal decision can be executed with a maximum complexity of $\mathcal{O}(L)$, enabling practical online implementation of the strategy.
\end{remark}
\vspace{-3mm}
\begin{remark}
An offline iterative algorithm, specified in Algorithm~\ref{Algorithm_off}, computes the average system throughput $\eta^*$, the unique solution to $\Omega(\eta^*)=\eta^*{\tau_s}$. The algorithm uses numerical accuracy $\epsilon$.
The convergence of sequence $\{\eta_m\}$, $m=1,\ldots,\infty$, is ensured by the Lipschitz continuity condition detailed in~\cite[Proposition 1.2.3]{Berts1999}, with an offline computational complexity of $\mathcal{O}(\log_2 \epsilon)$.
\end{remark}
\vspace{-1mm}
\vspace{-2mm}
\begin{algorithm}[h!]
	\caption{Iterative algorithm for $\eta^*$}\label{Algorithm_off}
 	\renewcommand{\algorithmicrequire}{\textbf{Input:}}
	\renewcommand{\algorithmicensure}{\textbf{Output:}}
 \begin{algorithmic}[1]
 \vspace{-1mm}
	\REQUIRE{$\epsilon$, $\eta_0=1$, $m=0$, $\Delta=1$, $\beta_1=\frac{\max\limits_{i=1,...,I}q_i}{\int_{0}^{\infty}\log_2(1+x)d F_{s}(x)}$}
	\WHILE{$\Delta\ge \epsilon$}
 \STATE 
$\Delta\leftarrow \Omega(\eta_m)-\eta_m{\tau_s}$
 
	\STATE	update $\eta_{m+1}\leftarrow\eta_m+\beta_2\cdot \Delta$,		
		where step-size $\beta_2$ satisfies $\epsilon\le \beta_2\le (2-\epsilon)/\big(\tau_s+\beta_1\big)$.
	\STATE	$m\leftarrow m+1$	
 \ENDWHILE
\STATE $\eta^*\leftarrow \eta_{m}$
 \end{algorithmic}
\end{algorithm}
\vspace{-1mm}

\begin{figure*}[t]
	\centering
	\subfloat[Average throughput vs $P_t$.]{
		\label{fig:comparison1}
		\includegraphics[width=0.32\textwidth]{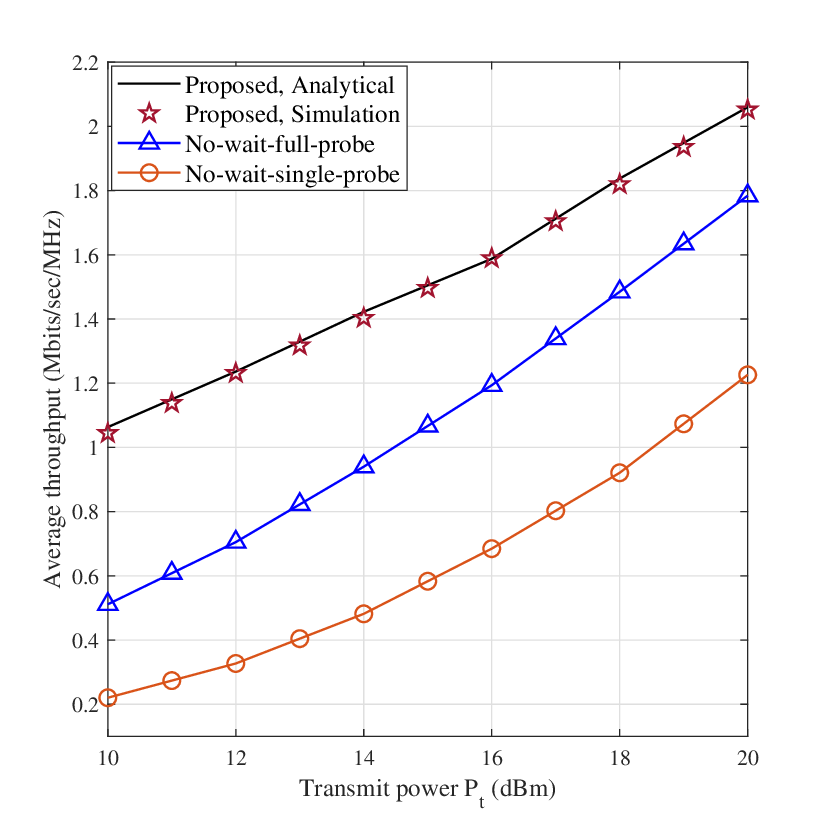}} 
	\subfloat[Average throughput vs size $Q$. ]{
		\label{fig:comparison2}
		\includegraphics[width=0.32\textwidth]{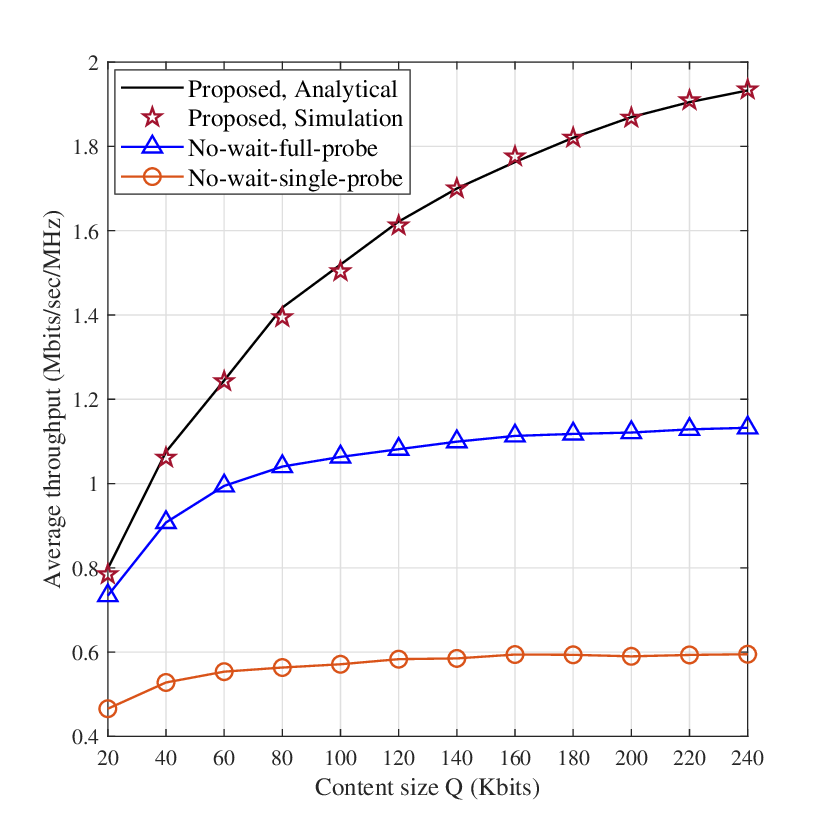}} 
  \subfloat[Average throughput vs interval $\tau_s$.]{
		\label{fig:comparison3}
		\includegraphics[width=0.32\textwidth]{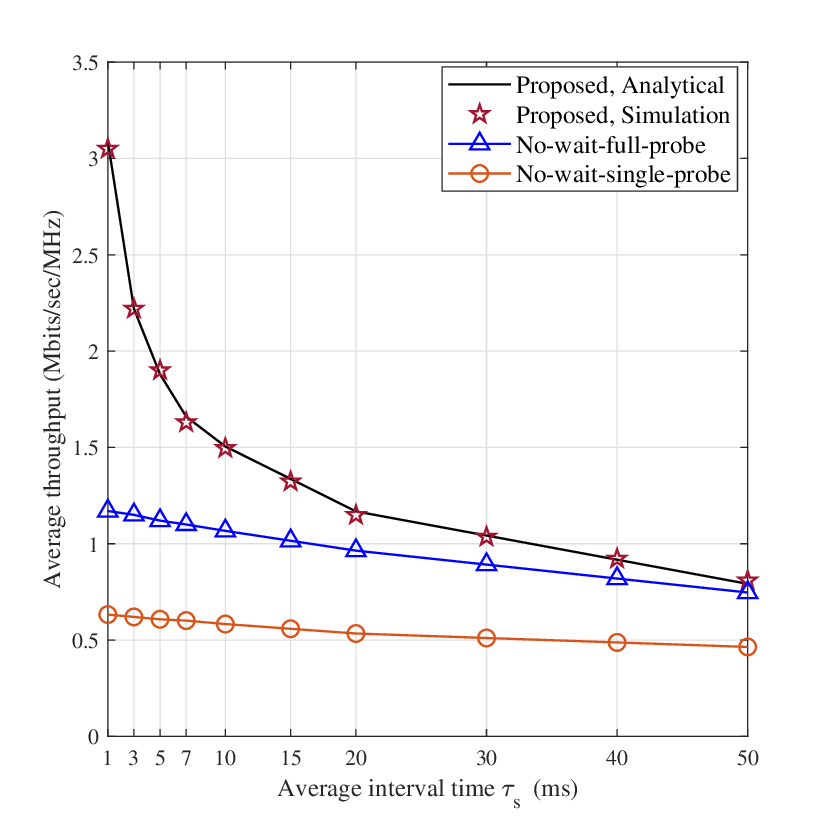}}
     \caption{Validation and performance comparison with alternative strategies.
     }
	\label{fig:comparison}
 \vspace{-6mm}
\end{figure*}

\vspace{-3mm}
\section{Numerical Results}\label{s:num}
\vspace{-1mm}
We study a network with a BS coverage radius $R=200$\,m and $L=5$ cooperative nodes positioned evenly on a circle with radius $50$\,m. The network has a path-loss exponent $\alpha=3$, a reference path-loss $\beta_0=-30$\,dB at $1$\,m, and a noise floor $N_0=-80$\,dBm, over a $1$\,MHz bandwidth. The system manages $I=8$ uniformly sized content items, each with a skewness parameter $\zeta=1.5$. The BS cache holds up to $C_b=3$ files, with equal caching probabilities of $3/8$. Each cooperative node stores one file ($C_r=1$) with equal caching probabilities of $1/8$. The probing time per node is $\tau=0.1$\,ms, and the latency for BS fetching file from the server is $T_1=0.1$\,ms.



Fig.~\ref{fig:comparison} compares average system throughput against transmit power $P_t$, content size $Q$, and request interval $\tau_s$, validating our theoretical results. It compares analytical results from Algorithm~1 with simulation outcomes from JCPUS strategy trials using Theorem~2. The close match between these results confirms our theoretical accuracy. Additionally, Fig.~\ref{fig:comparison} benchmarks our strategy against two alternatives:
i) \textit{No-wait-full-probe}: The BS efficiently probes all nodes without delay, optimizing user scheduling with a blend of single-node DF relaying and multi-node cooperative delivery~\cite{tang2023}.
ii) \textit{No-wait-single-probe}: The BS promptly schedules users without delay by randomly probing a cooperative node, utilizing an advanced method combining node DF relaying and cache-aided delivery~\cite{tang2023}.
Fig.~\ref{fig:comparison1} evaluates average system throughput versus transmit power $P_t$ for $Q=100$\,kbits and $\tau_s=10$ ms, showing our strategy outperforms alternatives with gains of at least $15.4$\% over the no-wait-full-probe and $67.9$\% over the no-wait-single-probe strategies. Throughput increases with $P_t$, and our strategy maintains superiority across all power levels.



Fig.~\ref{fig:comparison2} shows average throughput versus content size $Q$ at $P_t=15$\,dBm and $\tau_s=10$\,ms. Our strategy consistently outperforms alternatives, with a $58.4$\% gain over the no-wait-full-probe and a $196.7$\% advantage over the no-wait-single-probe at $Q=160$\,kbits. Throughput gains increase with larger content sizes, underscoring enhanced performance from caching and time diversity benefits.
Fig.~\ref{fig:comparison3} displays average throughput versus content request interval $\tau_s$ at $P_t=15$\,dBm and $Q=100$\,kbits. Our strategy outshines alternatives, with a $31.6$\% energy efficiency gain over the no-wait-full-probe and a $139.3$\% throughput advantage over the no-wait-single-probe at $\tau_s=15$\,ms. Throughput decreases as $\tau_s$ lengthens, with our strategy's advantage narrowing for $\tau_s \geq 50$\,ms due to reduced time diversity benefits impacting opportunistic scheduling.

\vspace{-1mm}
\section{Conclusion}\label{s:con}
\vspace{-1mm}
This study developed a novel analytical framework for a cache-aided cooperative network, focusing on opportunistic scheduling through cooperative probing and cache-aided delivery. Using optimization theory of sequentially planned decisions (SPD), we derived an optimal Reward Expected Thresholds (RET)-based rule, which was implemented as a Joint Cooperative Probing and User Scheduling (JCPUS) strategy. Utilizing the time-invariant statistical properties of observation processes, we proposed an optimal JCPUS strategy to maximize the average system throughput, with an online complexity of $\mathcal{O}(L)$, where $L$ is the number of cooperative nodes. Furthermore, we introduced iterative algorithms for offline implementation, ensuring practicality of the strategy. Our strategy effectively leverages caching gain, spatial diversity, and time diversities, resulting in enhanced performance. 

\end{document}